\documentclass[11pt,a4paper]{article}
\usepackage{amsfonts,amssymb,amsmath,amsthm,cite}
\usepackage{graphicx}
\usepackage{slashed}

\evensidemargin=\oddsidemargin
\newcommand{\MM}{\mathcal{M}}
\newcommand{\ii}{\mathrm{i}}
\newcommand{\nn}{\mathbf{n}}
\newcommand{\Tr}{{\mathrm{Tr}\,}}
\newcommand{\dd}{\mathrm{d}}

\newcommand{\be}{\begin{equation}}
\newcommand{\ee}{\end{equation}}
\newcommand{\bea}{\begin{eqnarray}}
\newcommand{\eea}{\end{eqnarray}}

\newtheorem{assumption}{Assumption}[section]
\newtheorem{theorem}[assumption]{Theorem}

\begin{document}
\title{Heat Kernel, Spectral Functions and Anomalies in Weyl Semimetals}
\author{A.~V.~Ivanov$^*$, M.~A.~Kurkov$^\dag$ and D.~V.~Vassilevich$^\ddag$\\
$^*$St. Petersburg Department of Steklov Mathematical Institute of
RAS,\\ 27 Fontanka, St. Petersburg 191023, Russia\\
$^*$Leonhard Euler International Mathematical Institute,\\ 10 Pesochnaya nab., St. Petersburg 197022, Russia\\
E-mail: regul1@mail.ru\\
$^\dag$ Dipartimento di Fisica ``E. Pancini'', Universit\`a di Napoli Federico II,\\ Complesso Universitario di Monte S. Angelo Edificio 6, via Cintia, 80126 Napoli, Italy.\\
$^\dag$ INFN-Sezione di Napoli, Complesso Universitario di Monte S. Angelo Edificio 6, \\ via Cintia, 80126 Napoli, Italy. \\
E-mail: max.kurkov@gmail.com \\
$^\ddag$ Center of Mathematics, Computation and Cognition, Universidade Federal do ABC\\
 09210-580, Santo Andr\'e, SP, Brazil\\
 E-mail: dvassil@gmail.com}
\date{\empty}
\maketitle
\begin{abstract}
Being motivated by applications to the physics of Weyl semimetals we study spectral geometry of Dirac operator with an abelian gauge field and an  {axial} vector field. We impose chiral bag boundary conditions with variable chiral phase $\theta$ on the fermions. We establish main properties of the spectral functions which ensure applicability of the $\zeta$ function regularization and of the usual heat kernel formulae for chiral and parity anomalies. We develop computational methods, including a perturbation expansion for the heat kernel. We  show that the terms in both anomalies which include electromagnetic potential are independent of $\theta$.
\end{abstract}

\section{Introduction and the statement of results}\label{sec:int}
Weyl semimetals \cite{Armitage:2017cjs} are without doubt among the most exciting new materials. The spectrum of quasiparticles in these materials is described by a Dirac operator containing a constant axial vector field $b$. This field is responsible for the existence of topologically protected boundary states with quite unusual dispersion relations -- the so called Fermi arcs. In the massless case, the field $b$ can be removed by a chiral transformation at the expense of including an anomaly term in the action. One of the manifestations of such anomaly term is the chiral magnetic effect \cite{Fukushima:2008xe,Kharzeev:2013ffa} 
(see also \cite{Alekseev:1998ds,Zyuzin:2012tv}) consisting in the appearance of an electric current in the direction of external magnetic field.

The importance of boundaries for physics of Weyl semimetals is very well appreciated. There have been many quantum computations for these materials in the presence of boundaries, though almost all of them were done in a simplified model where all boundary contributions were taken into account through an effective $(2+1)$-dimensional model describing boundary states. Such models do not always provide a good approximation to full quantum effective action as has been demonstrated recently in \cite{Fialkovsky:2019rum,Kurkov:2020jet}. There were a few calculations done for full quantum systems \cite{Gorbar:2015wya,Sitenko:2016nsm,Beneventano:2019qxm} for specific geometries with boundaries or compact dimensions.

Quantum anomalies are very important objects of Quantum Field Theory. They can be (almost entirely) expressed through integrals of local densities and thus carry information about generic geometries and field configurations. Though the anomalies do not describe the quantum effective action completely, the chiral anomaly tells us how this action changes when the field $b$ is turned off by means of a chiral transformation. Such an important quantity as the Hall conductivity at zero external magnetic field is defined by the so-called parity anomaly. The aim of the present work is to study these two anomalies on 4-dimensional manifolds with boundaries by means of the heat kernel methods (see \cite{Vassilevich:2003xt} for a physicist oriented introduction). As we shall see in a moment, a consistent formulation of this problem requires chiral bag boundary conditions with a varying chiral phase $\theta$ on $\partial\MM$. Very little is known about the spectral geometry of Dirac operators with these conditions. (We will give a short literature survey below in this Section.) Therefore, before starting calculations, we shall have to establish basic properties of main spectral functions of the Dirac operator with these boundary conditions. To be able to use the full power of spectral theory of elliptic operators, we will work in Euclidean signature, assuming a kind of Wick rotation in quantum effective action.

In this paper, we work on a smooth 4-dimensional Riemannian manifold $\MM$ with a smooth boundary $\partial\MM$. This manifold is equipped with a spin structure, a $U(1)$ gauge connection, and an axial vector field $b$. We shall \emph{not} assume that $b$ is constant in most of this work. The corresponding vector bundle over $\MM$ will be denoted by $\mathcal{S}$. Locally, the Dirac operator acting on sections of $\mathcal{S}$ has the form
\begin{equation}
\slashed{D}=\ii \gamma^\mu (D_\mu + \gamma_5 b_\mu),\label{Dir}
\end{equation}
where $\gamma^\mu$ denotes Dirac gamma matrices, $\gamma^\mu\gamma^\nu+\gamma^\nu\gamma^\mu=2g^{\mu\nu}$. The covariant derivative $D_\mu$ contains an electromagnetic potential $A$ and a spin connection $w$. In components, $D_\mu=\partial_\mu+ \tfrac 18 [\gamma_\rho,\gamma_\sigma]w_\mu^{\rho\sigma}+\ii A_\mu$. $D$ is compatible with the Clifford structure, $D_\mu\gamma^\nu=\gamma^\nu D_\mu$. The chirality matrix $\gamma_5$ is defined as
\begin{equation}
\gamma_5=\tfrac 1{4!}\epsilon^{\mu\nu\rho\sigma}\gamma_\mu\gamma_\nu\gamma_\rho\gamma_\sigma\,,\label{g5}
\end{equation}
where $\epsilon^{\mu\nu\rho\sigma}$ is the totally antisymmetric Levi-Civita tensor.

In the applications to condensed matter problems, the Dirac operator contains the Fermi velocity $v_F$. The operator (\ref{Dir}) is written under the assumption $v_F=1$. The dependence of quantum effective action on $v_F$ can be restored by using a suitable  modification of procedure \cite{Fialkovsky:2019rum}.

The classical action
\begin{equation}
S=\int_\MM \dd^4x \sqrt{g} \psi^\dag(x)\slashed{D}\psi(x) \label{clac}
\end{equation}
is invariant under local chiral transformations
\begin{equation}
\slashed{D}\to \slashed{D}_\phi= e^{\gamma_5\phi}\slashed{D}e^{\gamma_5\phi},\qquad \psi\to \psi_\phi= e^{-\gamma_5\phi}\psi,\label{chitrafo}
\end{equation}
where $\phi$ is a smooth function. The transformation of $\slashed{D}$ in (\ref{chitrafo}) is equivalent to a gradient transformation of $b$,
\begin{equation}
b_\mu\to b_\mu + \partial_\mu\phi \label{chib}.
\end{equation}
Thus, if $b$ is a constant vector or a gradient of a scalar function, it can be completely removed from the action. If $\partial\MM$ is empty, the whole effect is the very well known anomaly term. If there is a boundary, the situation is much more complicated. One has to find a suitable set of boundary conditions which  {respects} chiral symmetry and compute boundary contributions to the anomaly for such boundary conditions.

A minimal set of local boundary conditions which is closed under chiral transformations is the so called chiral bag boundary conditions
\begin{equation}
\Pi_-(\theta)\psi\vert_{\partial\MM}=0, \qquad \Pi_-(\theta)=\tfrac 12 \bigl( 1-\ii \varepsilon_\alpha \gamma_5 e^{\gamma_5\theta}\gamma^{\nn}\bigr).\label{Pith}
\end{equation}
Here $\gamma^\nn=\nn_\mu \gamma^\mu$ where $\nn$ is an inward pointing unit normal vector to $\partial\MM$, and  {$\theta$ is a smooth function on $\partial\mathcal{M}$}. $\varepsilon_\alpha$ is a real parameter, $\varepsilon_\alpha=\pm 1$, which is constant on each of the connected components $\partial\MM_\alpha$ of the boundary $\partial\MM$. These conditions originally appeared in the context of bag model of hadrons \cite{Rho:1983bh} (see also \cite{Hrasko:1983sj}), thus the name. With these conditions, the normal component of fermion current $\psi^\dag \gamma^\nn \psi$ vanishes identically on $\partial\MM$ making the Dirac operator symmetric. One can easily verify, that
\begin{equation}
\Pi_-(\theta-2\phi)\psi_\phi\vert_{\partial\MM}=0,\label{bcpsiphi}
\end{equation}
so that the conditions (\ref{Pith}) indeed form an orbit of the group of chiral transformations. Various spectral functions of the Dirac operator with chiral bag boundary conditions were discussed in \cite{Wipf:1994dy,Durr:1996im,Esposito:2002vz,Beneventano:2003hv,Esposito:2005dn,Gilkey:2005qm,Kirchberg:2006wu} -- but only for $\theta=const$, which is not sufficient to incorporate local chiral transformations.
 
In what follows, we shall mostly consider an infinitesimal version of the chiral transformations
\begin{equation}
\delta_\phi b_\mu=\partial_\mu (\delta\phi),\qquad \delta_\phi\theta=-2(\delta\phi),\label{infchi}
\end{equation}
where $\delta\phi$ is an (infinitesimal) parameter. It is understood that the second equation in (\ref{infchi}) should have contained the restriction of $\delta\phi$ to the boundary. We shall never write this restriction explicitly. 

Our purpose is to analyse the anomalies arising in the effective action obtained by integrating out the fermion field $\psi$. 
We use the $\zeta$-function regularization of the effective action \cite{Dowker:1975tf,Hawking:1976ja} (see \cite{Fursaev:2011zz} for a pedagogical introduction). The $\zeta$ function of $\slashed{D}$ is defined as
\begin{equation}
\zeta(s,\slashed{D})=\sum_{\lambda>0}\lambda^{-s} + e^{-\ii \pi s}\sum_{\lambda<0}(-\lambda)^{-s}, \label{zetas}
\end{equation}
where $\lambda$ denotes the eigenvalues of $\slashed{D}$. The sums are convergent if $\Re s$ is sufficiently large. The $\zeta$ function can be extended to the whole complex plane as a meromorphic function. These properties are valid for strongly elliptic boundary conditions. We postpone the discussion of this point until Section \ref{sec:main}. The regularized determinant of $\slashed{D}$ and the regularized effective action are defined as
\begin{equation}
W_s=-[\ln\det  {(\slashed{D})}]_s\equiv \Gamma(s)\zeta(s,\slashed{D}).\label{Ws}
\end{equation}
The regularization is removed in the limit $s\to 0$. Next, we represent $\zeta(s,\mathcal{D})$ as a sum of two expressions, one being symmetric and the other -- antisymmetric with respect to the inversion $\mathcal{D}\to - \mathcal{D}$. The symmetric part is expressed through $\zeta(s/2,\slashed{D}^2)$ while the antisymmetric is defined through the $\eta$ function
\begin{equation}
\eta(s,\slashed{D})=\sum_{\lambda>0}\lambda^{-s} -\sum_{\lambda<0}(-\lambda)^{-s}. \label{etas}
\end{equation} 
Thus, we have
\begin{equation}
W_s=\tfrac 12 \Gamma(s)\left[ \bigl(1+e^{-\ii \pi s}\bigr) \zeta(s/2,\slashed{D}^2)+
\bigl(1-e^{-\ii \pi s}\bigr)\eta(s,\slashed{D})\right].\label{Ws1}
\end{equation}
Let us assume that both $\zeta(s/2,\slashed{D}^2)$ and $\eta(s,\slashed{D})$ are regular at $s=0$. By expanding $W_s$ near $s=0$ we obtain
\begin{equation}
W_s=\left( \tfrac 1s +\mbox{f.t.}\right)\zeta(0,\slashed{D}^2)+\tfrac 12 \zeta'(0,\slashed{D}^2)+\frac{\ii\pi}2 \eta(0,\slashed{D})+\mathcal{O}(s),\label{Ws2}
\end{equation}
where f.t. denotes some finite terms and $\zeta'(0,\slashed{D}^2):=\partial_s|_{s=0} \zeta (s,\slashed{D}^2)$. The expression (\ref{Ws2}) is divergent at $s\to0$. The pole term has to be removed by some sort of renormalization procedure. There remains a finite renormalization ambiguity which consists in keeping a finite term proportional to $\zeta(0,\slashed{D}^2)$. Usually, such terms do not influence quantum anomalies. Let us assume that this is true in our case as well. Therefore, without losing any essential information we may write the renormalized effective action as
\begin{equation}
W=\tfrac 12 \zeta'(0,\slashed{D}^2) +\frac{\ii\pi}2 \eta(0,\slashed{D}).\label{RSformula}
\end{equation}
This is a well known formula. The derivation was also rather standard. We presented it here in full detail to highlight the assumptions which had to be made. These assumptions will be justified by Theorem \ref{Th1}, see below.

Calculation of $W$ for generic choice of $\MM$, $A$, and $b$ is a hopeless task. However, the anomalies associated to $W$ are computable. The first one is the chiral or Adler--Bell--Jackiw anomaly
\begin{equation}
\mathcal{A}=\tfrac 12 \delta_\phi\, \zeta'(0,\slashed{D}^2).\label{chian}
\end{equation}
The second one is the whole second term in (\ref{RSformula})
\begin{equation}
W_{\rm odd}=\frac{\ii\pi}2 \eta(0,\slashed{D}),\label{paran}
\end{equation}
which, by somewhat stretching the terminology, we shall call the parity anomaly. This anomaly was introduced \cite{Niemi:1983rq,Redlich:1983dv,AlvarezGaume:1984nf} to measure the violation of spatial reflection symmetry in quantum theories on manifolds without boundaries. It also measures the spectral asymmetry of Dirac operator. 

Very little is known about these anomalies in the presence of boundaries. The chiral anomaly was calculated for a constant $\theta$ in two dimensions \cite{Wipf:1994dy} and for $\theta=0$ in four dimensions \cite{Marachevsky:2003zb}. The parity anomaly with $\theta=0$ in four dimensions was calculated in \cite{Kurkov:2017cdz,Kurkov:2018pjw}. 

The anomalies (\ref{chian}) and (\ref{paran}) contain both local and global contributions. Local contributions are smooth universal functionals of external fields. Global contributions are in general non-smooth since they are related to zero modes of the Dirac operator. To separate the local parts we shall assume that $\slashed{D}$ \emph{does not have zero modes}. 

Our strategy is as follows. Since the spectral theory of Dirac operators subject to chiral bag boundary conditions with varying chiral parameter $\theta$ is an essentially uncharted area first we have to prove basic properties of the spectral problem and spectral functions which have been announced above. Such properties are usually taken for granted in the physics literature even though there are numerous counterexamples of boundary conditions for which they do not hold. Thus, some attention is required to put quantum field theory computations of firm grounds. General properties of our spectral problem are summarized in the following

\begin{theorem}\label{Th1}
Let $\slashed{D}$ be a Dirac operator on a smooth four-dimensional manifold $\MM$ with a smooth boundary $\partial\MM$ subject to chiral bag boundary conditions as defined above. \\
(a) Let $Q\in C^\infty (\mathrm{End}(\mathcal{S}))$ be a smooth endomorphism (a matrix-valued function) with a compact support. Then, there are full asymptotic expansions at $t\to +0$
\begin{eqnarray}
&& \Tr \bigl( Q e^{-t\slashed{D}^2}\bigr)\simeq \sum_{n=0}^\infty t^{(n-4)/2}a_n(Q,\slashed{D}^2),\label{asan}\\
&& \Tr \bigl( Q\slashed{D} e^{-t\slashed{D}^2}\bigr)\simeq \sum_{n=0}^\infty t^{(n-5)/2}a_n^\eta(Q,\slashed{D}),\label{asaneta}
\end{eqnarray}
with locally computable coefficients $a_n$ and $a_n^\eta$.\\
Let in addition $\MM$ be compact. Assume that $\slashed{D}$ has not zero eigenvalues. Then we have the following assertions.\\
(b) The $\eta$ function $\eta(s,\slashed{D})$ is regular at $s=0$.\\
(c) The $\zeta$ function $\zeta(0,\slashed{D}^2)$ is invariant under chiral transformations (\ref{infchi}).\\
(d) The chiral anomaly (\ref{chian}) reads
\begin{equation}
\mathcal{A}=-2a_4(\gamma_5(\delta\phi),\slashed{D}^2).\label{chianT}
\end{equation}
\noindent
(e) Let $\delta\slashed{D}$ be the variation of $\slashed{D}$ induced by a smooth (infinitesimal) variation of $A$ and $b$. Then
\begin{equation}
\delta\eta(0,\slashed{D})=-\frac 2{\sqrt{\pi}}\, a_3(\delta\slashed{D},\slashed{D}^2).
\label{deleta}
\end{equation}
\end{theorem} 

Next, we proceed with actual calculation of the anomalies by using the heat kernel expansion. The heat kernel coefficients which define the anomalies are the bulk and boundary integrals of local polynomials in the fields entering our problem and their derivatives having appropriate symmetry properties and correct mass dimension. (These two requirements will be explained in detail in Section \ref{sec:WZ} below.) The coefficients of these polynomials are functions of $\theta$. Our task, therefore, is to define all allowed polynomials and to compute the corresponding functions. Here, we face two problems as compared to simpler types of boundary conditions.

The first problem is that the number of allowed structures is enormous. To keep combinatorial complexity under the control we have to impose some restrictions. First of all, we assume that the Riemannian metric on $\MM$ is flat and that the boundary $\partial\MM$ is totally geodesic (i.e., the extrinsic curvature of $\partial\MM$ vanishes). The first assumption is rather natural if we keep in mind the condensed matter applications even though a non-flat effective metric may appear due to deformations or due to position dependence of the Fermi velocity. The second assumption means that $\partial\MM$ is sufficiently smooth so that the inverse curvature radii are small compared to other characteristic scales of the model. Even with these restrictions, the number of relevant invariants is very large. We shall explicitly compute  {all the} terms containing the electromagnetic potential $A$ since they are most interesting from the physical point of view. We shall also present a universal method which allows to obtain closed expressions for all remaining invariants and compute some of them.

The second problem is that many standard methods of calculation of the heat kernel coefficients \cite{GilkeyNew} do not work for chiral bag boundary conditions. For example, the (very important) Lemma 3.1.6 \cite{GilkeyNew} is not applicable since the boundary conditions cannot be presented in the required (product) form. However, the Wess--Zumino consistency conditions which are consequences of the Lie algebra composition law for the generators of anomalous symmetries are still very useful. These conditions allow to reduce considerably the number of unknown functions by using relatively simple algebraic methods. The rest of the computations have to be done with the help of perturbative expansions for the heat kernel. Our main technical  {tool} will be  {a version of} the heat kernel derived in \cite{Beneventano:2003hv} for cylindrical geometries and a Dyson-type expansion obtained below.

From the physical point of view, our principal result means that the most relevant parts of both anomalies are stable against variations of $\theta$. To proceed, we need to introduce some geometric notations. As above, the subscript (superscript) $\nn$ will denote normal components of vectors and tensors on the boundary while Latin indices $i,j,k$ will denote the tangential components. The Levi-Civita tensor on the boundary is defined as
\begin{equation}
\epsilon^{ijk}\equiv \epsilon^{\nn ijk}.\label{LCb}
\end{equation}
Let semicolon denote the covariant derivative computed with the Christoffel symbol $\Gamma^{\mu}_{\nu\rho}$ corresponding to the bulk metric $g$, e.g. $b_{\mu;\nu}=\partial_\nu b_\mu -\Gamma^\rho_{\mu\nu}b_\rho$. The same quantity corresponding to induced metric $h$ on $\partial\MM$ will be denoted by double dot.

\begin{theorem}[Partial stability of the anomalies]\label{Th2}
Let $\slashed{D}$ be a Dirac operator (\ref{Dir}) without zero eigenvalues subject to chiral bag boundary conditions on a compact manifold $\MM$ as described above. Then the terms containing electromagnetic potential in chiral and parity anomalies do not depend on $\theta$ and read 
\begin{eqnarray}
&&\mathcal{A}(A)=\frac 1{16\pi^2}\int_\MM \dd^4 x \sqrt{g}(\delta\phi)\epsilon^{\mu\nu\rho\sigma}F_{\mu\nu}F_{\rho\sigma},\\
&&W_{\mathrm{odd}}(A)=-\, \frac {\ii}{16\pi} \int_{\partial\MM}\dd^3 x\, {\sqrt{h}} \,\varepsilon_\alpha \, \epsilon^{\nn ijk} A_i A_{k:j}.
\end{eqnarray}
\end{theorem}

The absence of $\theta$-dependent terms which are quadratic in $A$ can be easily obtained on symmetry grounds, see Section \ref{sec:WZ}. Therefore, the main message of Theorem \ref{Th2} is the absence of terms which are linear in $A$. Such  {contributions} would generate linear (tadpole) terms in the effective action and mean an instability of Weyl semimetals against emission of photons. No such effects have been detected so far\footnote{There is a related effect of emission of photons by quasiparticles in Weyl semimetals predicted in \cite{Andrianov:2020lub}. This effect is very tiny and exists already at the tree level.}. Our present work gives a theoretical explanation for the absence of instabilities in anomalous parts of the effective action.

One is of course curious whether the stability statement is valid for the whole anomalies, including also the terms independent of $A$. The answer is negative, and to show this we calculate a couple of terms in the anomalies.

This paper is organized as follows. In the next section we establish regularity of the $\eta$ function and derive variational properties of $\zeta$ function. Section \ref{sec:WZ} studies general form of the heat kernel coefficients appearing in the anomalies and restrictions imposed on these coefficients by various symmetry properties. Perturbation theory for the heat kernel is derived in Section \ref{sec:per} and used for actual calculations in Section \ref{sec:cal}. Concluding remarks can be found in the last section.

\section{Main properties of the spectral functions}\label{sec:main}
Since we are going to consider the spectral problem for $\slashed{D}^2$, which is a second-order differential operator, we need a second boundary condition in addition to (\ref{Pith}). It reads 
\begin{equation}
\Pi_-(\theta)\slashed{D}\psi\vert_{\partial\MM}=0. \label{2ndBC}
\end{equation}
The heat kernel $K_{\slashed{D}^2}(x,y;t)$ of $\slashed{D}^2$ satisfies the heat equation
\begin{equation}
\bigl( \partial_t +\slashed{D}_x^2\bigr) K_{\slashed{D}^2}(x,y;t)=0,\label{heq}
\end{equation}
the initial condition
\begin{equation}
K_{\slashed{D}^2}(x,y;0)=\delta(x,y)\label{Kin}
\end{equation}
with $\delta(x,y)$ being the kernel of unit operator, and the boundary conditions
\begin{equation}
\Pi_-(\theta(x))K_{\slashed{D}^2}(x,y;t)=0,\qquad \Pi_-(\theta(x))\slashed{D}_x K_{\slashed{D}^2}(x,y;t)=0, \label{bchk}
\end{equation}
when $x\in \partial\MM$. If $\MM$ is non-compact, the boundary conditions (\ref{bchk}) must be supplemented by suitable falloff conditions at infinity.

Both $\slashed{D}$ and $\slashed{D}^2$ are elliptic differential (rather than pseudo-differential) operators. The strong ellipticity of spectral problems for $\slashed{D}$ and $\slashed{D}^2$ with chiral bag boundary conditions and a constant $\theta$ was demonstrated in \cite{Beneventano:2003hv}. Since the proof is local with respect to the points on the boundary, it is also valid for a varying $\theta$ without changes. Besides, the boundary conditions (\ref{Pith}) and (\ref{2ndBC}) are local. Thus, according to general theory (see \cite[Theorem 1.4.5]{GilkeyNew}), the heat kernel expansion is local and does not contain logarithmic term, which is exactly assertion (a) of Theorem \ref{Th1}. Locality means that the coefficients $a_n$ and $a_n^\eta$  can be expressed as bulk and boundary integrals of local invariant polynomials depending on the fields entering $\slashed{D}$ and boundary conditions. We will discuss the structure of heat kernel coefficients in more detail in the next section. Our results  {for the $\zeta$ and $\eta$ functions and for anomalies} are formulated on a compact $\MM$. Thus, the assumption of $Q$ having a compact support is redundant. However, the local integral formulae for heat kernel coefficients are valid in the non-compact case as well.  {In particular, for the sake of convenience, we shall analyze the relevant heat kernel coefficients on a non-compact manifold in Section~\ref{sec:per}.} There, the assumption regarding the support of $Q$ becomes essential.

There are the following integral representations for generalized $\zeta$ and $\eta$ functions containing a smooth endomorphism $Q$
\begin{eqnarray}
&&\zeta(s,\slashed{D}^2;Q)=\Tr \bigl( Q (\slashed{D}^2)^{-s}\bigr)=\frac 1{\Gamma(s)}\int_0^{+\infty} dt\, t^{s-1}\Tr \bigl( Qe^{-t\slashed{D}^2}\bigr),\label{zetaint}\\
&&\eta (s,\slashed{D};Q)=\Tr \bigl( Q \slashed{D}(\slashed{D}^2)^{-\frac{s+1}2}\bigr)=\frac 1{\Gamma\left(\frac{s+1}2\right)}\int_0^{+\infty} dt\, t^{\frac{s-1}2}\Tr \bigl( Q\slashed{D} e^{-t\slashed{D}^2}\bigr). \label{etaint}
\end{eqnarray}
These integral representations together with asymptotic expansions (\ref{asan}) and (\ref{asaneta}) define the pole structure of $\zeta(s,\slashed{D}^2;Q)$ and $\eta (s,\slashed{D};Q)$. They have only simple poles with
\begin{eqnarray}
&&a_k(Q,\slashed{D}^2)=\mbox{Res}_{s=\frac{4-k}2} \left( \Gamma(s)\zeta(s,\slashed{D}^2;Q)\right), \label{akres}\\
&&a_k^\eta (Q,\slashed{D})=\frac 12 \mbox{Res}_{s={4-k}} \left( \Gamma\left(\frac{s+1}2\right)\eta(s,\slashed{D};Q)\right),\label{aetares}
\end{eqnarray}
(see e.g. \cite{Gilkey:2005qm}).

To analyse chiral variation of the heat trace we use a (modified) technical trick suggested by Gilkey and Kirsten \cite{Gilkey:2005qm} for boundary conditions with $\theta=const$. To generalize this trick for arbitrary $\theta(x)$ one should extend smoothly $\theta$ inside $\MM$. Then, one defines a new operator $P=e^{-\frac 12 \theta \gamma_5}\slashed{D}e^{\frac 12 \theta \gamma_5}$. The multiplication by $e^{\frac 12 \theta\gamma_5}$ changes the boundary conditions. Thus, for the operator $P$ the boundary condition is $\Pi_-(0)\psi|_{\partial\MM}=0$. This boundary condition does not depend on $\theta$. The heat kernel for $P^2$ reads 
\begin{equation}
K_{P^2}(x,y;t)=e^{-\frac 12 \theta(x) \gamma_5}K_{\slashed{D}^2}(x,y;t)e^{\frac 12 \theta (y)\gamma_5}.\label{hkP2}
\end{equation}
One can easily check that this kernel satisfies equations (\ref{heq}), (\ref{Kin}), and (\ref{bchk}) with $P$ instead of $\slashed{D}$ and $\theta=0$. For any smooth endomorphism $Q$, which commutes with $e^{\frac 12 \theta \gamma_5}$, we have
\begin{equation}
\Tr \bigl( Q\, e^{-tP^2}\bigr) = \Tr \bigl( Q\, e^{-t\slashed{D}^2}\bigr).
\end{equation}
One can write $P=e^{-\theta\gamma_5}\slashed{D}_{b\to B=b+\frac 12 \partial\theta}$. Under chiral transformations (\ref{infchi}), the field $B$ does not change. The boundary conditions also remain invariant. Thus, the whole dependence on chiral phase resides in the prefactor $e^{-\theta\gamma_5}$. Thus, the chiral variation of heat trace of $\slashed{D}^2$ can be written as
\begin{align}\nonumber
\delta_\phi\, \Tr \bigl(e^{-t\slashed{D}^2}\bigr) &=
\delta_\phi\, \Tr \bigl(e^{-tP^2}\bigr) =-2t\ \Tr \bigl(\delta_\phi P \cdot P \cdot e^{-tP^2}\bigr)=-4t\ \Tr \bigl((\delta\phi)\gamma_5 P^2 \cdot e^{-tP^2}\bigr)\nonumber\\
& =4t\partial_t\Tr \bigl((\delta\phi)\gamma_5 e^{-tP^2}\bigr)=4t\partial_t\Tr \bigl( (\delta\phi)\gamma_5 e^{-t\slashed{D}^2}\bigr).\label{varTr}
\end{align}
This equation determines chiral variations of main spectral functions.

In the case of constant $\theta$, one can prove \cite{Gilkey:2005qm} that some heat kernel coefficients do not depend on $\theta$. With a varying $\theta$, there are similar statements meaning chiral invariance of corresponding quantities. Let us expand the first and the last terms in the chain of equations (\ref{varTr}) in asymptotic series and equate the terms with identical powers of $t$. This yields 
\begin{equation*}
\delta_\phi\, a_n(1,\slashed{D}^2)=2(n-4)a_n((\delta\phi)\gamma_5,\slashed{D}^2).
\end{equation*}
By setting $n=4$ we obtain
\begin{equation}
\delta_\phi\, a_4(1,\slashed{D}^2)=0.\label{chia4}
\end{equation}
The coefficient $a_4(1,\slashed{D}^2)$ is usually related to the global scale anomaly. Thus, this anomaly is invariant under local chiral transformations  as in the case of manifold without boundary. By using (\ref{akres}) we get $a_4(1,\slashed{D}^2)=\zeta(0,\slashed{D}^2)$. Thus, formula (\ref{chia4}) yields assertion (c) of Theorem \ref{Th1}.

By similar arguments, we have
\begin{equation}
\delta_\phi\,\Tr \bigl( \slashed{D} e^{-t\slashed{D}^2}\bigr)=
2(1+2t\partial_t)\, \Tr \bigl( (\delta\phi)\gamma_5\slashed{D} e^{-t\slashed{D}^2}\bigr).
\end{equation}
An expansion of this equation in asymptotic series yields
\begin{equation*}
\delta_\phi\, a_n^\eta(1,\slashed{D})=2(n-4)a_n^\eta((\delta\phi)\gamma_5,\slashed{D}).
\end{equation*}
For $n=4$ one gets the following chiral invariance condition
\begin{equation}
\delta_\phi\, a_4^\eta (1,\slashed{D})=0.\label{chia4eta}
\end{equation}
One can connect by a smooth chiral transformation the operator $\slashed{D}$ with boundary conditions (\ref{Pith}) to a  {chirally} transformed operator $\slashed{D}_\phi$ with boundary conditions with a constant or even vanishing $\theta$. The coefficient $a_4^\eta$ does not change. For the transformed operator this $a_4^\eta(1,\slashed{D}_\phi)=0$, as was demonstrated by Gilkey and Kirsten \cite{Gilkey:2005qm}. Therefore, $a_4^\eta(1,\slashed{D})=0$ also for arbitrary non-constant $\theta$. By setting $k=4$ in (\ref{aetares}) we obtain
\begin{equation}
\mbox{Res}_{s=0}\eta(s,\slashed{D})=2\pi^{-1/2}a_4^\eta(1,\slashed{D}_\phi)=0.\label{regeta}
\end{equation}
Thus, the eta function $\eta(s,\slashed{D})$ is regular at $s=0$. This completes the demonstration of Theorem \ref{Th1}(b).

The methods for computing variations of spectral functions can be found in \cite{Atiyah:1980jh} and \cite{Gilkey:1984}.
To compute chiral variation of $\zeta(s,\slashed{D})$ we substitute (\ref{varTr}) in the integral representation (\ref{zetaint}), integrate by parts, assuming that $s$ is sufficiently large, and use (\ref{zetaint}) again to get the identity
\begin{equation}
\delta_\phi\, \zeta(s,\slashed{D}^2)=-4s \zeta(s,\slashed{D}^2;(\delta\phi)\gamma_5).
\end{equation}
Since $\zeta(s,\slashed{D}^2;(\delta\phi)\gamma_5)$ is regular at $s=0$, we obtain
\begin{equation}
\delta_\phi\, \zeta'(0,\slashed{D}^2)=-4 \zeta(0,\slashed{D}^2;(\delta\phi)\gamma_5)=-4a_4((\delta\phi)\gamma_5,\slashed{D}^2).\label{varzprime}
\end{equation}
This yields assertion (d) of Theorem \ref{Th1}.

The last assertion (e) of Theorem \ref{Th1} is obtained by using the general methods \cite{Gilkey:1984,Atiyah:1980jh}. There is nothing specific for our case in this respect. We remark that the formula (\ref{deleta}) is valid only if the number of zero modes does not change due to the variation. This is a part of our general assumption that $\slashed{D}$ does not have any zero modes in the relevant range of parameters.

\section{General form of the anomalies and the Wess--Zumino conditions}\label{sec:WZ}
Scaling dimension is an important notion which allows to restrict possible form of the heat kernel coefficients. Let us take a formal parameter $\Lambda$ and rescale the proper time $t\to \Lambda^{-2}t$. We say, that $t$ has a canonical dimension $-2$. Canonical dimensions will be denoted by square brackets, $[t]\equiv -2$. To keep the expression $e^{-t\slashed{D}^2}$ scaling invariant, we rescale $\slashed{D}\to \Lambda \slashed{D}$, so that $[\slashed{D}]=1$. This behaviour of the Dirac operator may be achieved by assigning $[b]=[A]=1$ and $[x^\mu]=-[\partial_\mu]=-1$. The boundary conditions remain invariant if $[\theta]=0$. The scaling dimensions defined above coincide with canonical mass dimensions of physical fields. The scaling dimensions of heat kernel coefficients can be obtained by comparing the dimensions on both sides of equation (\ref{asan}). Since $Q$ appears linearly, one can assign to it any dimension. Thus, we obtain
\begin{equation}
[a_n(Q,\slashed{D}^2)]=n-4+[Q].\label{anQdim}
\end{equation}
The heat kernel coefficients are given by sums of bulk and boundary integrals. Since $[\dd^k x]=-k$, the canonical dimension of bulk integrands is $n+[Q]$, while boundary integrands have dimension $n-1+[Q]$.

\subsection{Contributions to the chiral anomaly containing electromagnetic potential}\label{sec:el}
Apart from having correct scaling behaviour, the heat kernel coefficient $a_4(\gamma_5\delta\phi,\slashed{D}^2)$ has to be gauge invariant. The most general form of this coefficient reads
\begin{align}
a_4(\gamma_5\delta\phi,\slashed{D}^2)=&-\frac 1{32\pi^2}\int_\MM \dd^4 x \sqrt{g}(\delta\phi)\epsilon^{\mu\nu\rho\sigma}F_{\mu\nu}F_{\rho\sigma}\nonumber\\
\qquad &+\frac 1{16\pi^2} \int_{\partial\MM}\dd^3 x\sqrt{h}(\delta\phi) ( b_iF_{jk}\epsilon^{ijk}f_1(\theta)+b^iF_{\nn i}f_2(\theta) \nonumber\\
\qquad\qquad &\qquad\qquad\,\,\,+ F_{ij}f_3(\theta)_{:k}\epsilon^{ijk}+F^{\nn i} f_4(\theta)_{:i} + f_5(\theta)F^{\nn i}_{\ \ :i} \bigr).\label{a45}
\end{align}
Since $[\theta]=0$, the field $\theta$ can enter with any power, so that the right hand side of (\ref{a45}) is defined up to five arbitrary functions $f_a(\theta)$. All these functions vanish at $\theta=0$, see \cite{Marachevsky:2003zb}. The volume term is the celebrated Adler--Bell--Jackiw term. Due to the locality of the heat kernel expansion the volume term does not depend on $\theta$.

Chiral transformations (\ref{infchi}) form an abelian Lie algebra,
\begin{equation}
\delta_{\phi_1}\circ \delta_{\phi_2}-\delta_{\phi_2}\circ \delta_{\phi_1}=0.\label{WZ1}
\end{equation}
By applying this relation to $\zeta'(0,\slashed{D}^2)$ and using (\ref{chian}) and (\ref{chianT}) we obtain the Wess--Zumino consistency condition
\begin{equation}
\delta_{\phi_1}a_4(\delta\phi_2\gamma_5,\slashed{D}^2)-\delta_{\phi_2}a_4(\delta\phi_1\gamma_5,\slashed{D}^2)=0.\label{WZ2}
\end{equation}
This condition can be easily solved yielding
\begin{equation}
f_1(\theta)-2f'_3(\theta)=0,\qquad f_2(\theta)-2f'_4(\theta)=0.\label{WZ3}
\end{equation}
Thus, it remains to compute only three of five functions $f_a(\theta)$.

\subsection{Parity anomaly}\label{subsec:par}
Here we study $A$-dependent terms in the parity anomaly. Let us take a variation $\delta A_\mu$ of the gauge field $A$. Then, $\delta\slashed{D}=-\gamma^\mu \delta A_\mu$. Since $\eta(0,\slashed{D})$ is gauge invariant, the variation can be represented in the form
\begin{equation}
\int_{\partial\MM}\dd^3 x \sqrt{h} \left[ (\delta A_j) J^j + (\delta F_{\nn j}) K^j \right],
\end{equation}
where $J^j$ is a conserved current, $J^{j}_{\ \ :j}=0$, while $K^j$ is gauge invariant.
Therefore, we get
\begin{align}
a_3(\delta\slashed{D},\slashed{D}^2)=\int_{\partial\mathcal{M}} \dd^3 x\sqrt{h} \Big( (\delta A_i)& \epsilon^{ijk} (A_k G_0(\theta))_{:j}  +(\delta A_i) \epsilon^{ijk} (b_k G_1(\theta))_{:j} \nonumber\\
\quad\quad
&+b^i\, \big(\delta F_{n i}\big)\,G_2(\theta)  + \big(\delta F^{\nn i}\big)\,G_3(\theta)\, \theta_{:i}  \Big).\label{a3stru}
\end{align}
The functions $G_a(\theta)$, $a=0,1,2,3$, have to be determined. 

Second variation of any functional (local or nonlocal) has to be symmetric. Thus, we have for two consequent variations $\delta A^{(1)}$ and $\delta A^{(2)}$ of the gauge field
\begin{equation*}
0=\bigl(\delta_{A^{(2)}}\circ \delta_{A^{(1)}}-\delta_{A^{(1)}}\circ \delta_{A^{(2)}}\bigr) \eta(0,\slashed{D})=-\frac 2{\sqrt{\pi}} \int_{\partial\MM}\dd^3 x\, \sqrt{h}\epsilon^{ijk}(\delta A^{(1)}_i)(\delta A^{(2)}_k)\theta_{:j}G'_0(\theta).
\end{equation*} 
Since this condition has to hold for arbitrary variations and an arbitrary $\theta(x)$, we  conclude that $G'_0=0$. The value of $G_0$ at $\theta=0$ has been computed in \cite{Kurkov:2017cdz}. Hence, we have
\begin{equation}
G_0(\theta)=\frac {\varepsilon_\alpha}{8\pi^{3/2}}.\label{G0}
\end{equation}
The integration of the term with $G_0$ in variation (\ref{a3stru}) yields second assertion of Theorem \ref{Th2}.

In Section \ref{sec:par} it will be shown that the functions $G_1$, $G_2$, and $G_3$ vanish.

\paragraph{Contributions to the anomalies not involving $A_\mu$.} Let us briefly describe the contributions to anomalies which have not been included in the previous discussion. Since these terms do not contain $A$, they are not restricted by the gauge invariance. Thus, many structures are allowed. For example, the chiral anomaly depends on more that 20 independent functions of $\theta$. However, about a half of these function can be expressed through the rest by means of the Wess--Zumino conditions. All terms in both anomalies can be calculated at least in principle with the perturbation expansion from the next section, though at higher orders the calculations become very time consuming. We present explicitly just a single boundary term in each the heat kernel coefficients corresponding to the anomalies
\begin{eqnarray}
&&a_4(\gamma_5\,\delta\phi\, \slashed{D}^2) = \int_{\partial\mathcal{M}} \dd^3 x\sqrt{h} \,f_6(\theta) \,\delta\phi_{;\nn\nn\nn}  + \cdots,\label{a4b}\\
&&a_3(\ii \gamma^{\mu}\gamma^5 (\delta b_{\mu}),\slashed{D}^2) =  \int_{\partial\mathcal{M}} \dd^3 z\, \sqrt{h}\,G_4(\theta)\, (\delta b_{\nn})_{;\nn\nn} + \cdots.  \label{etaa3}
\end{eqnarray} 
The volume terms in $a_4$ are fixed as in \cite{Marachevsky:2003zb}. $f_6$ and $G_4$ will be computed in Section \ref{sec:more}.

\section{Perturbation theory for the heat kernel}\label{sec:per}
As we have seen above, the relevant heat kernel coefficients are defined up to several unknown local functions of the boundary chiral angle $\theta$. To compute these functions, it is enough to consider the simplest one-boundary geometry $\mathbb{R}^3\times \mathbb{R}_+$. In this case we can introduce the following decomposition $x=(x^{\|},x^\nn)\in\mathbb{R}^3\times \mathbb{R}_+$, where $x^{\|}\in\mathbb{R}^3$ and $x^\nn\in\mathbb{R}_+$.

The transformation $\psi\to\gamma_5\psi$ maps $\slashed{D}\to-\slashed{D}$ and inverts the signs $\varepsilon_\alpha\to -\varepsilon_\alpha$ in the boundary conditions. Since $\zeta(s,\slashed{D}^2)$ is invariant under the reflection $\slashed{D}\to-\slashed{D}$, it also has to be invariant under $\varepsilon_\alpha\to -\varepsilon_\alpha$. Similarly, $\eta(s,\slashed{D})$ has to change sign under $\varepsilon_\alpha\to -\varepsilon_\alpha$. Thus, in the case with a single connected boundary component it is sufficient to consider a selected value of $\varepsilon$ (we take $\varepsilon=1$). Complete dependence of spectral functions on this parameter will be restored below by using the mentioned property of the spectral functions. 

We will use perturbation theory expansions in $A$, $b$, and derivatives of $\theta$. The unperturbed heat kernel $K_0(x,y;t)$ will correspond to $A=0$, $b=0$, and a constant $\theta$. Also, let us introduce 
\be
\xi = x^\nn - y^\nn,\quad \eta = x^\nn + y^\nn, \quad \sigma = \big|x^{\|}- y^{\|}\big|,
\ee
and
\be
c = \cosh{(\theta)}, \quad s = \sinh{(\theta)}, \quad \tau = \tanh{(\theta)}. \label{hypnot}
\ee

A useful expression for the heat kernel $K_0$ on a semi-infinite cylinder $\mathcal{N}\times \mathbb{R}_+$ with a compact $\mathcal{N}$ was derived in \cite{Beneventano:2003hv}. We take $\mathcal{N}=T^3$. In the limit of infinite radii, one obtains the following expression for $K_0(x,y;t)$ on $\mathbb{R}^3\times \mathbb{R}_+$
\be
{K}_0(x,y;t) = \mathcal{K}(x,y,t)\, \mathrm{Id}
 - \frac{1}{\tau}\,\partial_{\nn[x]}\mathcal{B}(x,y,t)\,\mathcal{P}_{+}
+ \partial_{j[x]}\mathcal{B}(x,y,t)\,\gamma_5\gamma^\nn\gamma^j\mathcal{P}_{+}, \label{K0}
\ee
with
\be
\mathcal{K}(x,y,t) = \frac{1}{(4\pi t)^{2}}  \Big(e^{-\frac{\xi^2 +\sigma^2}{4t}} - e^{-\frac{\eta^2+\sigma^2}{4t}} \Big), \label{Adef}
\ee
and
\be
\mathcal{B}(x,y,t) = -\frac{\tau{c}^{2}e^{-\frac{\eta^2 +\sigma^2}{4t}}}{4\pi^{2}t}\, 
\left( -\frac{1}{\eta}\, U\left(\frac{\sigma\tau}{\eta},\frac{t}{\eta^2 c^2}\right) \,+\left(\frac{\tau\,}{\sigma}\right)V\left(\frac{\sigma\tau}{\eta},\frac{t}{\eta^2 c^2}\right) \right),  \label{BU}
\ee
where $V(x,t)$ and $U(x,t)$ stand for Voigt profiles defined as
\be
U(u,t) = \frac{1}{\sqrt{4\pi t}}\int_{-\infty}^{+\infty}\dd y\,\, \frac{e^{-\frac{(u-y)^2}{4t}}}{y^2+1},
\quad V(u,t) = \frac{1}{\sqrt{4\pi t}}\int_{-\infty}^{+\infty}\dd y\,\, \frac{y\,e^{-\frac{(u-y)^2}{4t}}}{y^2+1}.
\ee
The Hermitian projectors are defined as
\be
\mathcal{P}_+ := \frac{\Pi_+^{\phantom{\dagger}}\Pi_+^{\dagger}}{\cosh^2{(\theta)}}, \quad \mathcal{P}_- := \frac{\Pi_-^{\dagger}\Pi_-^{\phantom{\dagger}}}{\cosh^2{(\theta)}}, \nonumber
\ee
where $\Pi_+:=\tfrac 12 \bigl( 1+\ii \gamma_5 e^{\gamma_5\theta}\gamma^{\nn}\bigr)$.

It can be also checked directly that the kernel (\ref{K0}) satisfies the heat equation, the initial condition, and the boundary conditions.

\subsection{Variation of $\theta$} \label{sec:vt}
To make the formulae easier to read we introduce a notation for the right action of $\slashed{D}$ on a matrix-valued kernel $\kappa(x,z)$,
\be
   \kappa(x,z)^{\dagger} \overset{\leftarrow}{\slashed{D}}(z):=(\slashed{D}(z)\kappa(x,z))^{\dagger}, \label{rightD}
\ee
where $\dagger$ denotes Hermitian conjugation of the matrices.

Let $K_{\slashed{D}^2,\theta} (x,y;t)$ and $K_{\slashed{D}^2,\theta+\Delta\theta} (x,y;t)$ be the heat kernels for the operator $\slashed{D}^2$ with the boundary conditions characterized by chiral angles $\theta$ and $\theta+\Delta\theta$, respectively. Since the symbol of the operator does not change, we do not include $\slashed{D}^2$ in the notations of the heat kernel.

We start with a sequence of obvious identities
\begin{eqnarray}
&&K_{\theta+\Delta\theta} (x,y;t)=K_\theta(x,y;t) + \int_{\mathcal{M}} \dd^4 z\,\sqrt{g}\,\left(\delta(x,z) K_{\theta+\Delta\theta}(z,y;t) - K_\theta (x,z;t)\delta(z,y)\right)\nonumber\\
&&\phantom{K_{\theta+\Delta\theta} (x,y;t)} =K_\theta(x,y;t)+\int_{\mathcal{M}} \dd^4 z\,\sqrt{g}\,\left(K_\theta(x,z;t-w) K_{\theta+\Delta\theta}(z,y;w)\right)\big|_{w = 0}^{w = t}
 \nonumber \\
&&\phantom{K_{\theta+\Delta\theta} (x,y;t)} 
=K_\theta(x,y;t)+\int_0^t \dd w\int_{\mathcal{M}} \dd^4 z\,\sqrt{g}\,\left(-\left(\partial_{t} K_\theta(x,z;t-w)\right) K_{\theta+\Delta\theta}(z,y;w)\right.\nonumber\\
&&\phantom{K_{\theta+\Delta\theta} (x,y;t)}
\qquad\qquad\qquad\qquad\qquad\qquad \left.
  + K_\theta (x,z;t-w)\partial_{w}K_{\theta+\Delta\theta}(z,y;w)\right) \nonumber \\ 
&&\phantom{K_{\theta+\Delta\theta} (x,y;t)} 
=K_\theta(x,y;t)+\int_0^t \dd w\int_{\mathcal{M}} \dd^4 z\,\sqrt{g}\,\left( \slashed{D}^2(x)  K_\theta(x,z;t-w) K_{\theta+\Delta\theta}(z,y;w)\right.\nonumber\\
&&\phantom{K_{\theta+\Delta\theta} (x,y;t)}
\qquad\qquad\qquad\qquad\qquad\qquad \left.
  - K_\theta(x,z;t-w)\slashed{D}^2(z)K_{\theta+\Delta\theta}(z,y;w)\right). 
\end{eqnarray}
The integration by parts yields
\begin{eqnarray}
&&K_{\theta+\Delta\theta} (x,y;t)=K_\theta(x,y;t) + \int_0^t \dd w\int_{\partial\mathcal{M}} \dd^3 z\, \sqrt{h}\Big( K_\theta(x,z;t-w) \overset{\leftarrow}{\slashed{D}}(z)
     \ii \gamma^\nn \, K_{\theta+\Delta\theta}(z,y;w) \nonumber\\
 &&\qquad\qquad\qquad\qquad\qquad\qquad\qquad\qquad\qquad    +K_\theta (x,z;t-w) 
     \ii \gamma^\nn  {\slashed{D}}(z) K_{\theta+\Delta\theta}(z,y;w)\Big). \label{interme1}
\end{eqnarray}
Since in the last line $z\in\partial\MM$, one can use the boundary conditions and rewrite factors in the form
\begin{eqnarray}
&&K_\theta(x,z;t-w) \overset{\leftarrow}{\slashed{D}}(z) = K_\theta(x,z;t-w) \overset{\leftarrow}{\slashed{D}}(z)\Pi_{+}^{\dagger}(\theta), \nonumber\\
&&K_\theta(x,z;t-w)  = K_\theta(x,z;t-w)\Pi_{+}^{\dagger}(\theta(z)), \nonumber\\
&& {\slashed{D}}(z) K_{\theta+\Delta\theta}(z,y;w) = \Pi_{+}(\theta(z) + \Delta\theta(z)) {\slashed{D}}(z) K_{\theta+\Delta\theta}(z,y;w), \nonumber\\
 &&K_{\theta+\Delta\theta}(z,y;w) = \Pi_{+}(\theta(z)+\Delta\theta(z))K_{\theta+\Delta\theta}(z,y;w), \label{BCS}
\end{eqnarray}
together with the identity
\be
\Pi_{+}^{\dagger}(\theta) \gamma^\nn \Pi_{+}(\theta+\Delta\theta) = \Pi_{+}^{\dagger}(\theta) \gamma^\nn (\Delta\Pi_+)\Pi_{+}(\theta+\Delta\theta),
\ee
where $\Delta\Pi_+:=\Pi_+(\theta+\Delta\theta)-\Pi_+(\theta)$. 
\be
K_{\theta + \Delta\theta}(x,y;t) = K_{\theta}(x,y;t) 
+  \int_0^t \dd w\int_{\partial\mathcal{M}} \dd^3 z\, \sqrt{h}\,\,K_{\theta}(x,z,t-w)\,\mathcal{P}(z)\, K_{\theta + \Delta\theta}(z,y;w), \label{mastereq}
\ee
with
\be
\mathcal{P}(z) := \overset{\leftarrow}{\slashed{D}}(z)  \ii \gamma^\nn (\Delta\Pi_+) + \ii\gamma^\nn (\Delta\Pi_+)   {\slashed{D}}(z) . \label{Pdef}
\ee

By taking an infinitesimal $\Delta\theta=\delta\theta$ and linearizing the equation above, we arrive at the following equation for the variation of the heat kernel
\begin{eqnarray}
&&\delta_{\theta}K_{\slashed{D}^2,\theta}(x,y;t) =  \frac{1}{2}\int_0^t \dd w \int_{\partial\mathcal{M}} \dd^3 z \,\sqrt{h}(z)\,K_{\slashed{D}^2,\theta}(x,z; t - w )
\nonumber\\
&&\qquad\qquad\qquad\qquad\qquad\qquad \times \Big( \overset{\leftarrow}{\slashed{D}}(z)\,\ii \gamma^n \gamma_5\, \delta\theta(z) +\ii \gamma^n \gamma_5\, \delta\theta(z) \,  {\slashed{D}}(z)\Big) K_{\slashed{D}^2,\theta}(z,y;w).  \label{App2}
\end{eqnarray}
\subsection{Variations of $A$ and $b$}\label{sec:vAb}
In this subsection we give smooth localized perturbation to $A$ and $b$. The Dirac operator changes $\slashed{D}\to\slashed{D}+\Delta\slashed{D}$. We assume that $\theta$ remains unperturbed and introduce the corresponding heat kernels by $K_{\slashed{D}^2}(x,y;t)$ and $K_{(\slashed{D}+\Delta\slashed{D})^2}(x,y;t)$, respectively.

By acting as in the previous subsection, we write
\begin{align}
K_{(\slashed{D}+\Delta\slashed{D})^2}(x,y;t)=K_{\slashed{D}^2}(x,y;t)+\int_0^t& \dd w\int_{\mathcal{M}} \dd^4 z\,\sqrt{g}
\left( \slashed{D}^2(x)  K_{\slashed{D}^2}(x,z;t-w) K_{(\slashed{D}+\Delta\slashed{D})^2}(z,y,w)\right.\nonumber\\
&\left.
- K_{\slashed{D}^2}(x,z;t-w) \big(\slashed{D}(z) + \Delta{\slashed{D}}(z)\big)^2K_{(\slashed{D}+\Delta\slashed{D})^2}(z,y;w)\right).\nonumber
\end{align}
After integrating by parts twice and using the boundary conditions we arrive at a Dyson-type equation
\be
 K_{(\slashed{D} + \Delta\slashed{D})^2}(x,y;t) =  K_{\slashed{D}^2}(x,y;t) +  \int_0^t \dd w\int_{\mathcal{M}} \dd^4 z\, \sqrt{g}\,\, K_{\slashed{D}^2}(x,z;t-w)\, P(z)\, 
  K_{(\slashed{D} + \Delta\slashed{D})^2}(z,y;w), \label{mastereqD}
\ee
where
\begin{equation}
P(z) = - \Big(\overset{\leftarrow}{\slashed{D}}(z)\,\Delta\slashed{D}(z) + \Delta\slashed{D}(z)\,\slashed{D}(z)\Big)-\big(\Delta\slashed{D}(z)\big)^2 
 . \label{PdefD}
\end{equation}

The variations of $\theta$, $A$, and $b$ can be repeated as many times as needed. At the end of the day, one arrives at an expansion which can be represented symbolically as $K=K_0+K_0 \hat P K_0 + K_0 \hat P K_0 \hat P K_0+\dots$, where $\hat P$ is either $P$ or $\mathcal{P}$. Repeated integrations over the proper time parameter $t$ and over either $\MM$ or $\partial\MM$ are understood. The orders of this expansion correspond to powers of $A$, $b$ or derivatives of $\theta$. Higher order derivatives of these fields are taken into account through an expansion of the kernel $K_0$, as one will see below. Each coefficient in the heat kernel expansion requires a finite order of the perturbation series.

\section{Calculation of the anomalies}\label{sec:cal}
\subsection{Proof of the partial stability}\label{sec:par}
To prove Theorem \ref{Th2} it remains to show that the functions $f_1$, $f_2$, $f_5$, $G_1$, $G_2$, and $G_3$ vanish. All these functions are associated with the structures which are linear in the field $A_\mu$. There is a general (though yet non-rigorous) argument which shows that such structures cannot appear in the heat kernel expansion. Indeed, let us introduce an antihermitian connection $\mathbf{A}=\ii A$ and antihermitian gamma matrices $\tilde\gamma^\mu=\ii \gamma^\mu$. Then the imaginary unit $\ii$ disappears from the Dirac operator and from the boundary conditions. It does not reappear in the trace rules for $\tilde{\gamma}$.  Since the heat expansion is an expansion over the symbols of $\slashed{D}$ and of the boundary operator, the imaginary unit cannot enter the coefficients. If, after computing the traces over spinor indices we return to the original field $A$, we observe that all expressions which are linear in $A$ acquire   {pure} imaginary coefficients. Thus, they have  {pure} imaginary values on the configurations, which we consider here. Since the heat kernel expansion of a self-adjoint operator has to be real, we conclude that the coefficient in front of such expressions vanish. 

It appeared quite hard to give precise mathematical meaning to the construction presented above on arbitrary manifold. However, on $\MM=\mathbb{R}^3\times\mathbb{R}_+$ the redefinitions described in the previous paragraph completely remove the imaginary unit from the unperturbed $K_0$ and from the operators $P$ and $\mathcal{P}$. Hence, the imaginary unit cannot appear at any order of the perturbations series. Since to define the unknown function in heat kernel coefficients it is sufficient to perform calculations on this simple background, we conclude that linear terms in $A$ are not allowed in the anomalies. This completes the proof of Theorem \ref{Th2}.

To illustrate the proof, we outline the main steps of perturbative calculations.

We start with the parity anomaly. To evaluate the functions $G_1(\theta)$ and $G_2(\theta)$ we need to compute
\begin{equation}
\mathrm{Tr}\,\Big[ (\delta\slashed{D}) e^{-t\slashed{D}^2} \Big] \label{1320}
\end{equation}
on $\MM=\mathbb{R}^3\times \mathbb{R}_+$, where $\delta\slashed{D}=-\gamma^\mu A_\mu$, while $e^{-t\slashed{D}^2}$ has to be expanded to the linear order in $b$ according to the equation (\ref{mastereqD}). The relevant expression reads
\begin{eqnarray}
&&\int_{\mathcal{M}}\dd^4 x\,\big( \delta A_{\mu}(x)\big)\int_0^t \dd w\int_{\mathcal{M}} \dd^4 z\, \mathrm{tr}\,\Big[ \gamma^{\mu} K_{0}(x,z;t-w)\,  \Big(\overset{\leftarrow}{\slashed{D}}_0(z)\,(- \ii   \gamma_5\, \gamma^{\nu}b_{\nu}(z))\nonumber\\
&&\qquad\qquad\qquad\qquad\qquad\qquad\qquad\qquad\qquad\qquad + (- \ii   \gamma_5\, \gamma^{\nu}b_{\nu}(z))\,\slashed{D}_0(z)\Big)\, 
  K_{0}(z,x;w) \Big]. \nonumber
\end{eqnarray}
By counting the $\gamma$ matrices, one immediately sees that the expression above is  {pure} imaginary and thus has to vanish. We conclude that
\begin{equation}
G_1(\theta)=G_2(\theta)=0.\label{G1G2}
\end{equation}

The expression with $G_3$ in (\ref{a3stru}) depends on the derivative of $\theta$. Thus, we need to take $\delta\slashed{D}=-\gamma^\mu A_\mu$ as above, while $e^{-t\slashed{D}^2}$ has to be expanded to the linear order in $\delta\theta$ by using (\ref{App2}). One arrives at
\begin{eqnarray}
&&- \frac{1}{2}\int_{\mathcal{M}}\dd^4 x\, \delta A_{\mu}(x)\int_0^t \dd w \int_{\partial\mathcal{M}} \dd^3 z \, \delta\theta(z)\,
\mathrm{tr}\,\big[\gamma^{\mu}\,
K_{0}(x,z; t - w )\Big( \overset{\leftarrow}{\slashed{D}}(z)\,\ii \gamma^\nn \gamma_5\,
\nonumber\\
&&\qquad\qquad\qquad\qquad\qquad\qquad\qquad\qquad\qquad\qquad\qquad\qquad\qquad
+\ii \gamma^\nn \gamma_5\, {\slashed{D}}(z)\Big) K_{0}(z,x;w)\big].\nonumber
\end{eqnarray}
This expression is  {pure} imaginary, therefore, we conclude that
\begin{equation}
G_3(\theta)=0.\label{G3}
\end{equation}

Now, let us consider the chiral anomaly. The term with $f_5$ requires the first order of the perturbation series for the heat kernel. Thus, the calculation goes exactly as above. The terms with $f_1$ and $f_2$ contain both $A$ and $b$ and thus require the second order of perturbation series in $\Delta\slashed{D}$:
\begin{eqnarray}
&& \int_0^t \dd w\int_0^w \dd q\int_{\mathcal{M}} \dd^4 z_1\int_{\mathcal{M}} \dd^4 z_2\int_{\mathcal{M}} \dd^4 x\,\mathrm{tr}\, \big[ \gamma_5\,\delta\phi(x)\, K_{0}(x,z_1;t-w)\nonumber\\
&&\qquad\qquad  \times\Big(\overset{\leftarrow}{\slashed{D}}_0(z_1)\,\Delta\slashed{D}(z_1) + \Delta\slashed{D}(z_1)\,\slashed{D}_0(z_1)\Big)
  K_{0}(z_1,z_2,w-q) \nonumber\\
&&\qquad\qquad \times\Big(\overset{\leftarrow}{\slashed{D}}_0(z_2)\,\Delta\slashed{D}(z_2) + \Delta\slashed{D}(z_2)\,\slashed{D}_0(z_2)\Big)\, K_{0}(z_2,x;q)\big].\nonumber
\end{eqnarray} 
One can easily check that all cross terms in the above expression, containing both $A$ and $b$, are {pure} imaginary after taking the trace over spinor indices. Therefore, they must vanish after summing up all individual contributions.
\subsection{More terms in the anomalies}\label{sec:more}
Here, we compute the functions $f_6(\theta)$ and $G_4(\theta)$ defined in equations (\ref{a4b}) and (\ref{etaa3}), respectively. The term (\ref{a4b}) does not depend on $A$ and $b$ and does not contain derivatives of $\theta$. Therefore, it can be calculated by using the zero order term
\be
\mathrm{Tr} \,\Big(\gamma_5\,\delta\phi\, e^{-t\slashed{D}_0^2} \Big) = \int_{\mathcal{M}} \dd^4 x\, \mathrm{tr}\,\big( \gamma_5\, \delta \phi(x)\, K_0(x,x,t)\big)\label{103}
\ee
in power series of $t$. The only non-vanishing trace is $\mathrm{tr}\, \big(\gamma_5\,\mathcal{P}_+\big) =2\tau$. At coinciding arguments $x=y$ the function $\mathcal{B}(x, y, t)$ does not depend on tangential coordinates. Let us introduce the notation $\mathcal{C}(x^\nn,t):=\mathcal{B}(x, x, t)$. The expression (\ref{103}) reads 
\be
 \int_{\mathbb{R}^3}\dd^3 x^{\|}\, \delta\phi(x^{\|},0)\,\mathcal{C}(x^\nn,t)
+  \int_{\mathbb{R}^3\times {\mathbb{R}_+}}\dd^4 x\, \partial_\nn\delta\phi(x^{\|},x^\nn)\,\mathcal{C}(x^\nn,t), \label{stepi1a}
\ee
where we have integrated by parts over $x^\nn$. The function $\mathcal{C}(x^\nn,t)$ decays exponentially fast at $x^\nn\to\infty$ for any positive $t$. To construct an asymptotic expansion at $t\to+0$ according to the conventional wisdom we introduce a rescaled coordinate $u$ as $x^\nn = u\cdot\sqrt{t}$. Then, 
$ {\mathcal{C}(x^\nn,t)=t^{-3/2}\mathcal{C}(u,1)}$. The integrals in (\ref{stepi1a}) become
\begin{equation}
\frac{1}{t^{\frac{3}{2}}}\int_{\mathbb{R}^3}\dd^3 x^{\|}\,\delta\phi(x^{\|},0)\,\mathcal{C}(0,1)+\frac{1}{t} \int_{\mathbb{R}^3}\dd^3 x^{\|}\int_0^{+\infty}\dd u\,  \partial_\nn\delta\phi\bigl(x^{\|},ut^{\frac 12}\bigr)
\mathcal{C}(u,1).\label{1338}
\end{equation}
We expand $\delta\phi$ in a Taylor series around $x^\nn=0$. This yields
\begin{equation}
a_k (\gamma_5\delta\phi ,\slashed{D}_0^2)=  c_k(\theta)\int_{\mathbb{R}^3} \dd^3 x^{\|}\,\partial_{\nn}^{k-1}\delta\phi (x^{\|},0),
\end{equation}
where for $k\geq 2$ we have
\begin{equation}
c_k(\theta)=\frac{1}{(k-2)!}\int_0^{+\infty} \dd u\,u^{k-2}\mathcal{C}(u,1).
\end{equation}
The integration can be performed with the use of Mathematica. We conclude that
\be
f_6(\theta) = c_4(\theta)=\frac{1}{16\pi^2}\, \coth{(\theta)}\, (\theta\, \coth{(\theta)} - 1). \label{nnnAns}
\ee

The calculation of $G_4$ goes exactly the same way. One obtains
\be
G_4(\theta) = -\frac{\varepsilon_\alpha c_3(\theta)}{\sinh{(\theta)}}=-\frac{\varepsilon_\alpha}{16\pi^{\frac{3}{2}}}\, \frac{\cosh{(\theta)}}{1+\cosh{(\theta)}}.
\ee
Here we restored the dependence of parity anomaly on $\varepsilon_\alpha$ according to the rule formulated at the beginning of Section \ref{sec:per}.

We see that in general both anomalies contain terms depending on $\theta$.

\section{Conclusions}\label{sec:con}
Being motivated by applications to Weyl semimetals we studied the heat kernel expansion, the $\eta$ and $\zeta$ functions, and anomalies for Dirac operators containing an abelian gauge field and an axial vector field subject to boundary conditions with arbitrary chiral phase $\theta$. We established main properties of the spectral functions (like the regularity of $\eta(s,\slashed{D})$ at $s=0$, for example) which justify the use of $\zeta$ regularization and usual formulae for the anomalies. We have analysed the restrictions imposed by symmetries on the heat kernel coefficients. A Dyson-type perturbation theory for the heat kernel has been derived and used for calculation of the anomalies. We computed all anomalous contributions containing electromagnetic field. 

The method of pertubative expansion of the heat kernel is about the same important as our other results. By using this method, one can compute for example the boundary terms in conformal anomaly which has a lot of applications. We mention the boundary central charges in Conformal Field Theories \cite{Herzog:2015ioa,Fursaev:2015wpa,Solodukhin:2015eca} anomaly induced boundary currents \cite{Chu:2018ksb}.

We are planning to study physical effects following from our calculations. There is, however, a problem to be resolved before doing this. One has to understand the Wick rotation to Minkowski signature, which is quite a non-trivial task in the presence of chiral phase $\theta$.

\paragraph{Acknowledgements.} The work of D.V.V. was supported in parts by the S\~ao Paulo Research Foundation (FAPESP), project 2016/03319-6, and by the grant 305594/2019-2 of CNPq.
The work of A.V.I. is supported by the grant in the form of subsidies from the Federal budget for state support of creation and development of world-class research centers, including international mathematical centers and world-class research centers that perform research and development on the priorities of scientific and technological development. The agreement is between MES and PDMI RAS from "8" November 2019  {number}
075-15-2019-1620. Also, A.V.I. is a winner of the Young Russian Mathematician award and would like to thank its sponsors and jury.

\bibliographystyle{fullsort}
\bibliography{graphene,parity}

\providecommand{\href}[2]{#2}\begingroup\raggedright\begin{thebibliography}{10}

\bibitem{Armitage:2017cjs}
N.~P. Armitage, E.~J. Mele, and A.~Vishwanath, ``{Weyl and Dirac Semimetals in
  Three Dimensional Solids},'' {\em Rev. Mod. Phys.} {\bf 90} (2018), no.~1,
  015001,
\href{http://www.arXiv.org/abs/1705.01111}{{\tt 1705.01111}}.

\bibitem{Fukushima:2008xe}
K.~Fukushima, D.~E. Kharzeev, and H.~J. Warringa, ``{The Chiral Magnetic
  Effect},'' {\em Phys. Rev. D} {\bf 78} (2008) 074033,
  \href{http://www.arXiv.org/abs/0808.3382}{{\tt 0808.3382}}.

\bibitem{Kharzeev:2013ffa}
D.~E. Kharzeev, ``{The Chiral Magnetic Effect and Anomaly-Induced Transport},''
  {\em Prog. Part. Nucl. Phys.} {\bf 75} (2014) 133--151,
  \href{http://www.arXiv.org/abs/1312.3348}{{\tt 1312.3348}}.

\bibitem{Alekseev:1998ds}
A.~Y. Alekseev, V.~V. Cheianov, and J.~Fr{\"o}hlich, ``{Universality of
  transport properties in equilibrium, Goldstone theorem and chiral anomaly},''
  {\em Phys. Rev. Lett.} {\bf 81} (1998) 3503--3506,
  \href{http://www.arXiv.org/abs/cond-mat/9803346}{{\tt cond-mat/9803346}}.

\bibitem{Zyuzin:2012tv}
A.~A. Zyuzin and A.~A. Burkov, ``{Topological response in Weyl semimetals and
  the chiral anomaly},'' {\em Phys. Rev. B} {\bf 86} (2012) 115133,
  \href{http://www.arXiv.org/abs/1206.1868}{{\tt 1206.1868}}.

\bibitem{Fialkovsky:2019rum}
I.~Fialkovsky, M.~Kurkov, and D.~Vassilevich, ``{Quantum Dirac fermions in a
  half-space and their interaction with an electromagnetic field},'' {\em Phys.
  Rev.} {\bf D100} (2019), no.~4, 045026,
\href{http://www.arXiv.org/abs/1906.06704}{{\tt 1906.06704}}.

\bibitem{Kurkov:2020jet}
M.~Kurkov and D.~Vassilevich, ``{How many surface modes does one see on the
  boundary of a Dirac material?},'' {\em Phys. Rev. Lett.} {\bf 124} (2020),
  no.~17, 176802,
\href{http://www.arXiv.org/abs/2002.06721}{{\tt 2002.06721}}.

\bibitem{Gorbar:2015wya}
E.~V. Gorbar, V.~A. Miransky, I.~A. Shovkovy, and P.~O. Sukhachov, ``{Chiral
  separation and chiral magnetic effects in a slab: The role of boundaries},''
  {\em Phys. Rev. B} {\bf 92} (2015), no.~24, 245440,
  \href{http://www.arXiv.org/abs/1509.06769}{{\tt 1509.06769}}.

\bibitem{Sitenko:2016nsm}
Y.~A. Sitenko, ``{On the chiral separation effect in a slab},'' {\em EPL} {\bf
  114} (2016), no.~6, 61001, \href{http://www.arXiv.org/abs/1603.09268}{{\tt
  1603.09268}}.

\bibitem{Beneventano:2019qxm}
C.~G. Beneventano, M.~Nieto, and E.~M. Santangelo, ``{Chiral magnetic effect at
  finite temperature in a field-theoretic approach},'' {\em J. Phys. A} {\bf
  53} (2020), no.~46, 465401, \href{http://www.arXiv.org/abs/1910.11425}{{\tt
  1910.11425}}.

\bibitem{Vassilevich:2003xt}
D.~V. Vassilevich, ``{Heat kernel expansion: User's manual},'' {\em Phys.
  Rept.} {\bf 388} (2003) 279--360,
\href{http://www.arXiv.org/abs/hep-th/0306138}{{\tt hep-th/0306138}}.

\bibitem{Rho:1983bh}
M.~Rho, A.~S. Goldhaber, and G.~E. Brown, ``{Topological Soliton Bag Model for
  Baryons},'' {\em Phys. Rev. Lett.} {\bf 51} (1983)
747--750.

\bibitem{Hrasko:1983sj}
P.~Hrasko and J.~Balog, ``{The Fermion Boundary Condition and the $\theta$
  Angle in {QED} in Two-dimensions},'' {\em Nucl. Phys. B} {\bf 245} (1984)
  118--126.

\bibitem{Wipf:1994dy}
A.~Wipf and S.~D{\"{u}}rr, ``{Gauge theories in a bag},'' {\em Nucl. Phys. B}
  {\bf 443} (1995) 201--232,
  \href{http://www.arXiv.org/abs/hep-th/9412018}{{\tt hep-th/9412018}}.

\bibitem{Durr:1996im}
S.~D{\"{u}}rr and A.~Wipf, ``{Finite temperature Schwinger model with chirality
  breaking boundary conditions},'' {\em Annals Phys.} {\bf 255} (1997)
  333--361, \href{http://www.arXiv.org/abs/hep-th/9610241}{{\tt
  hep-th/9610241}}.

\bibitem{Esposito:2002vz}
G.~Esposito and K.~Kirsten, ``{Chiral bag boundary conditions on the ball},''
  {\em Phys. Rev.} {\bf D66} (2002) 085014,
\href{http://www.arXiv.org/abs/hep-th/0207109}{{\tt hep-th/0207109}}.

\bibitem{Beneventano:2003hv}
C.~G. Beneventano, P.~B. Gilkey, K.~Kirsten, and E.~M. Santangelo, ``{Strong
  ellipticity and spectral properties of chiral bag boundary conditions},''
  {\em J. Phys.} {\bf A36} (2003) 11533,
\href{http://www.arXiv.org/abs/hep-th/0306156}{{\tt hep-th/0306156}}.

\bibitem{Esposito:2005dn}
G.~Esposito, P.~Gilkey, and K.~Kirsten, ``{Heat kernel coefficients for chiral
  bag boundary conditions},'' {\em J. Phys.} {\bf A38} (2005) 2259--2276,
\href{http://www.arXiv.org/abs/math/0510156}{{\tt math/0510156}}.

\bibitem{Gilkey:2005qm}
P.~Gilkey and K.~Kirsten, ``{Stability theorems for chiral bag boundary
  conditions},'' {\em Lett. Math. Phys.} {\bf 73} (2005) 147--163,
\href{http://www.arXiv.org/abs/math/0510152}{{\tt math/0510152}}.

\bibitem{Kirchberg:2006wu}
A.~Kirchberg, K.~Kirsten, E.~M. Santangelo, and A.~Wipf, ``{Spectral asymmetry
  on the ball and asymptotics of the asymmetry kernel},'' {\em J. Phys.} {\bf
  A39} (2006) 9573--9589,
\href{http://www.arXiv.org/abs/hep-th/0605067}{{\tt hep-th/0605067}}.

\bibitem{Dowker:1975tf}
J.~S. Dowker and R.~Critchley, ``{Effective Lagrangian and Energy Momentum
  Tensor in de Sitter Space},'' {\em Phys. Rev. D} {\bf 13} (1976) 3224.

\bibitem{Hawking:1976ja}
S.~W. Hawking, ``{Zeta Function Regularization of Path Integrals in Curved
  Space-Time},'' {\em Commun. Math. Phys.} {\bf 55} (1977) 133.

\bibitem{Fursaev:2011zz}
D.~Fursaev and D.~Vassilevich, {\em {Operators, Geometry and Quanta}}.
\newblock Theoretical and Mathematical Physics. Springer, Berlin, Germany,
2011.
\newblock

\bibitem{Niemi:1983rq}
A.~J. Niemi and G.~W. Semenoff, ``{Axial Anomaly Induced Fermion
  Fractionization and Effective Gauge Theory Actions in Odd Dimensional
  Space-Times},'' {\em Phys. Rev. Lett.} {\bf 51} (1983)
2077.

\bibitem{Redlich:1983dv}
A.~N. Redlich, ``{Parity Violation and Gauge Noninvariance of the Effective
  Gauge Field Action in Three-Dimensions},'' {\em Phys. Rev.} {\bf D29} (1984)
2366--2374.

\bibitem{AlvarezGaume:1984nf}
L.~Alvarez-Gaume, S.~Della~Pietra, and G.~W. Moore, ``{Anomalies and Odd
  Dimensions},'' {\em Annals Phys.} {\bf 163} (1985)
288.

\bibitem{Marachevsky:2003zb}
V.~N. Marachevsky and D.~V. Vassilevich, ``{Chiral anomaly for local boundary
  conditions},'' {\em Nucl. Phys.} {\bf B677} (2004) 535--552,
\href{http://www.arXiv.org/abs/hep-th/0309019}{{\tt hep-th/0309019}}.

\bibitem{Kurkov:2017cdz}
M.~Kurkov and D.~Vassilevich, ``{Parity anomaly in four dimensions},'' {\em
  Phys. Rev.} {\bf D96} (2017), no.~2, 025011,
\href{http://www.arXiv.org/abs/1704.06736}{{\tt 1704.06736}}.

\bibitem{Kurkov:2018pjw}
M.~Kurkov and D.~Vassilevich, ``{Gravitational parity anomaly with and without
  boundaries},'' {\em JHEP} {\bf 03} (2018) 072,
\href{http://www.arXiv.org/abs/1801.02049}{{\tt 1801.02049}}.

\bibitem{GilkeyNew}
P.~B. Gilkey, {\em {Asymptotic formulae in spectral geometry}}.
\newblock CRC Press, Boca Raton, 2004.

\bibitem{Andrianov:2020lub}
A.~Andrianov, R.~Soldati, and D.~Vassilevich, ``{Emission of Photons by
  Quasiparticles in Weyl Semimetals},'' {\em Symmetry} {\bf 12} (2020), no.~5,
  869, \href{http://www.arXiv.org/abs/2003.13952}{{\tt 2003.13952}}.

\bibitem{Atiyah:1980jh}
M.~F. Atiyah, V.~K. Patodi, and I.~M. Singer, ``{Spectral asymmetry and
  Riemannian geometry. III},'' {\em Math. Proc. Cambridge Phil. Soc.} {\bf 79}
  (1976)
71--99.

\bibitem{Gilkey:1984}
P.~B. Gilkey, {\em {Invariance theory, the heat equation, and the Atiyah-Singer
  index theorem}}.
\newblock Publish or Perish, Wilmington, 1984.

\bibitem{Herzog:2015ioa}
C.~P. Herzog, K.-W. Huang, and K.~Jensen, ``{Universal Entanglement and
  Boundary Geometry in Conformal Field Theory},'' {\em JHEP} {\bf 01} (2016)
  162,
\href{http://www.arXiv.org/abs/1510.00021}{{\tt 1510.00021}}.

\bibitem{Fursaev:2015wpa}
D.~Fursaev, ``{Conformal anomalies of CFT’s with boundaries},'' {\em JHEP}
  {\bf 12} (2015) 112,
\href{http://www.arXiv.org/abs/1510.01427}{{\tt 1510.01427}}.

\bibitem{Solodukhin:2015eca}
S.~N. Solodukhin, ``{Boundary terms of conformal anomaly},'' {\em Phys. Lett.}
  {\bf B752} (2016) 131--134,
\href{http://www.arXiv.org/abs/1510.04566}{{\tt 1510.04566}}.

\bibitem{Chu:2018ksb}
C.-S. Chu and R.-X. Miao, ``{Anomaly Induced Transport in Boundary Quantum
  Field Theories},''
\href{http://www.arXiv.org/abs/1803.03068}{{\tt 1803.03068}}.

\end{thebibliography}\endgroup

\end{document}